\documentclass[a4paper]{jpconf}
\usepackage{graphicx}

\begin{document}
\title{DARWIN \\dark matter WIMP search with noble liquids}

\author{Laura Baudis, for the DARWIN consortium\footnote{The following institutions are members of the consortium (as of  December 2011): 
ETH~Z\"{u}rich, University of Z\"{u}rich (CH); 
Karlsruhe, Mainz, MPIK~Heidelberg, M\"{u}nster (DE); 
Nikhef (NL);
Subatech (FR); 
Weizmann Institute of Science (IL);
INFN (IT): Bologna, L'Aquila, LNGS, Milano, Milano Bicocca, Napoli, Padova, Pavia, Perugia, Torino; 
Associated groups: Columbia, Princeton, UCLA, Arizona State (USA).}}

\address{Physics Institute, University of Z\"urich, CH-8057 Z\"urich, Switzerland}

\ead{laura.baudis@physik.uzh.ch}

\begin{abstract}

DARWIN (dark matter wimp search with noble liquids) is a design study for a next-generation, multi-ton dark matter detector in Europe.  Liquid argon and/or liquid xenon  are the target media for the direct detection of dark matter candidates in the form of weakly interacting massive particles (WIMPs). Light and charge signals created by particle interactions in the active detector volume are observed via the time projection chamber technique.  DARWIN is to probe  the spin-independent, WIMP-nucleon cross section down 10$^{-48}$cm$^2$ and to measure WIMP-induced nuclear recoil spectra with high-statistics, should they be discovered by an existing or near-future experiment.  After a brief introduction,  I will describe the project,  selected R\&D topics,  expected backgrounds and the physics reach.

\end{abstract}

\section{Introduction}
\label{introduction}

Astrophysical observations show that 83\% of gravitating matter in our universe is non-luminous and non-baryonic. The dark matter might be in the form of neutral, weakly interacting, stable or long-lived elementary particles, so-called {\small WIMPs}, which have eluded direct observation so far. Produced in the early universe, WIMPs would naturally lead to the observed dark matter abundance \cite{lee77} and are predicted to exist in extensions to the Standard Model of particle physics \cite{feng10}. Galactic {\small WIMPs} may be detected via scatters off atomic nuclei in deep underground experiments \cite{goodman85}. Since expected signal rates are below one interaction per kilogram of target material and year and momentum transfers are around 10\,MeV - 100\,MeV \cite{jkg96, smith_lewin96},  large detector masses,  low energy thresholds and ultra-low backgrounds are essential experimental requirements to directly observe these hypothetical particles.   
Experiments using xenon \cite{xenon10,xenon100_prl105,xenon100_prl107,zeplin3} and argon \cite{warp} as a homogeneous detection medium in a time projection chamber  have reached sensitivities of  $\sim$10$^{-44}$cm$^2$ for the spin-independent scattering cross section on nucleons. While a factor of five improvement is predicted with data already in hand \cite{xenon100_prl107}, ton-scale experiments under commissioning \cite{ardm} or construction \cite{xenon1t_tdr}  will  likely probe the cross section region down to  $\sim$5$\times$10$^{-47}$cm$^2$.  
Notwithstading this remarkable leap in sensitivity, and assuming a local density and velocity distribution which are inferred from astronomical observations, significantly larger detectors are requisite to determine {\small WIMP} properties, such as its mass, scattering cross section and possibly spin \cite{complementarity}. 
To convincingly demonstrate the dark matter nature of a signal, a measurement of its interaction rate with multiple target materials is compulsory. 

\section{Technologies}
\label{technologies}

The {\small DARWIN} study \cite{darwin_2010,darwin_2011,darwin} is focused on a multi-ton liquid argon and/or xenon experiment rooted in the noble liquid time projection chamber  {\small (TPC)} technique.  The {\small TPCs} will record the prompt scintillation light\footnote{The peak photon emission from the transition of exited dimers into their dissociative ground state is centered around 128\,nm and 178\,nm in liquid argon and xenon, respectively \cite{aprile_book}.} created when a particle interacts in the active detector volume along with the few liberated electrons after they are drifted in a strong electric field\footnote{Typical drift fields are 0.5--1\,kV/cm, with electron velocities around $\sim$2\,mm/$\mu$s \cite{aprile_book}.} and extracted into the vapor phase residing above the liquid.  The prompt light signal will be observed by an array of photosensors immersed in the liquid, the electrons will be detected either directly, or indirectly via proportional scintillation in the gas phase with a second array of photosensors. The time difference between the prompt and delayed signals determines the $z$-position of an event, the spatial distribution of the delayed signal yields its $x-y$-position. The relative size of the charge and light signals, as well as their time structure will be used to distinguish nuclear recoils, as expected from WIMP scatters, from electronic recoils, which make the majority of the background. The spatial resolution allows to define an innermost, low-background volume and to reject fast neutrons, which -- in contrast to WIMPs -- tend to multiple scatter\footnote{The mean free path of  $\sim$MeV neutrons is in the range of tens of cm.}.

{\small DARWIN} will immensely benefit from the research and development, and from the construction and operation experience gained with  {\small XENON10} \cite{xenon10_instrument},  {\small XENON100} \cite{xenon100_instrument},  {\small XENON1T} \cite{xenon1t_tdr},  {\small WARP} \cite{warp}, ArDM \cite{ardm}, DarkSide \cite{darkside}, and much  of the ongoing work is carried out within the framework of these projects. Here I mention a few studies only, some of these are specific to {\small DARWIN}. Other work deals with the cryogenic, gas purification, circulation, storage and recovery systems; with the external water Cerenkov shield and its potential extension with a liquid scintillator (depending on the depth of the underground laboratory -- the Gran Sasso Laboratory and the Modane extension are under consideration); with material screening, selection and radon emanation measurements; with high-voltage systems, electrodes and field uniformity simulations; with low-noise, low-power electronics,  cables and connectors, trigger schemes, data acquisition and treatment;  with Monte Carlo simulations of the expected background noise, of the light collection efficiency and position reconstruction capability;  with the design of the time projection chamber, of the cryostat and of the calibration system. 

{\sl Light and charge response:} the light and charge yields of noble liquids when exposed to low-energy nuclear recoils (from neutron or potential dark matter interactions) or electronic recoils (from $\gamma$- and $\beta$-interactions) are studied by several groups participating in {\small DARWIN}. A new measurement of the relative scintillation efficiency ${\cal L}_\textnormal{\footnotesize eff}$ in liquid xenon \cite{leff}  shows an ${\cal L}_\textnormal{\footnotesize eff}$ behavior which is  slowly decreasing   with energy, with a non-zero value at 3\,keV nuclear recoil energy, the lowest measured point. A similar measurement is  ongoing for liquid argon~\cite{regenfus}. A measurement of the liquid xenon scintillation efficiency for electronic recoils down to 2.3\,keV is in progress ~\cite{manalaysay}. A  preliminary data analysis indicates that the scintillation yield falls with decreasing energy, as predicted by models of scintillation mechanisms in noble liquids \cite{szydagis11}. Nonetheless, the scintillation response at 2.3\,keV is observed to be non-zero, confirming that liquid xenon experiments will have a finite sensitivity at such low interaction energies. Measurements of the charge yields  of LAr and LXe within the same energy regime  are being planned. 

{\sl Signal readout:} the prompt scintillation light is to be observed either with conventional photomultiplier tubes {\small (PMTs)} which are low in radioactivity  \cite{screening} and built to withstand low temperatures and high pressures\footnote{One example is the Hamamatsu R11410/R11065 3''-tube for LXe/LAr, currently tested for its performance, long-term stability and radioactivity levels at several  {\small DARWIN} institutions.} or with a new, hybrid photodetector {\small (QUPID \cite{qupid})}, which has an extremely low  radioactivity content  ($<$1\,mBq/sensor for U/Th/K/Co) \cite{screening} and works both in liquid argon and xenon. 
The delayed signal can be observed directly, using detectors with single electron sensitivity and high spatial granularity (large electron multipliers \cite{lems}), or {\small CMOS} pixel detectors coupled to electron multipliers (GridPix \cite{gridpix}), or via proportional scintillation in the gas phase, using gaseous photomultipliers  without dead zones ({\small GPMs} \cite{gpms}), {\small  PMTs} or {\small QUPIDs}.

\section{Backgrounds, physics reach and timeline}
\label{goals}

{\small DARWIN} will be  an ``ultimate''   argon and/or xenon dark matter experiment, before the solar and atmospheric neutrinos become the main, possibly irreducible background. It will directly probe  {\small WIMP}-nucleon cross sections  down to  $\sim$10$^{-48}$cm$^2$. These cross sections are compatible with recent LHC results, should the dark matter particle  turn out to be the neutralino \cite{fowlie11,buchmueller11,strege11}. 
The external background from gammas, muons and neutrons and the background from detector construction materials will be diminished to negligible levels by external shields, the self-shielding of the noble 
liquids\footnote{The mean free path of 3\,MeV gammas is $\sim$9\,cm and $\sim$20\,cm in liquid xenon and argon, respectively.}, and the choice of fiducial volumes\footnote{The final choice of the size and target materials are part of the outcome of the study, a baseline scenario is 20\,t (10\,t)  total (fiducial) LAr/LXe mass.}. More difficult are intrinsic backgrounds from  $^{85}$Kr and $^{222}$Rn decays in  xenon and from $^{39}$Ar decays in argon. In xenon, the natural krypton concentration is to be reduced by cryogenic distillation to $<$1\,ppt and the radon level in the liquid is to be kept  $<$1\,$\mu$Bq/kg. Argon gas that is extracted from deep underground wells is depleted in the radioactive $^{39}$Ar  \cite{mei_prc10}.  Still, a background rejection by pulse-shape analysis of  $>$10$^8$ is required in the case of a liquid argon detector \cite{darwin_2010,darwin_2011}.

\begin{figure}[h!]
\includegraphics[width=8.0cm]{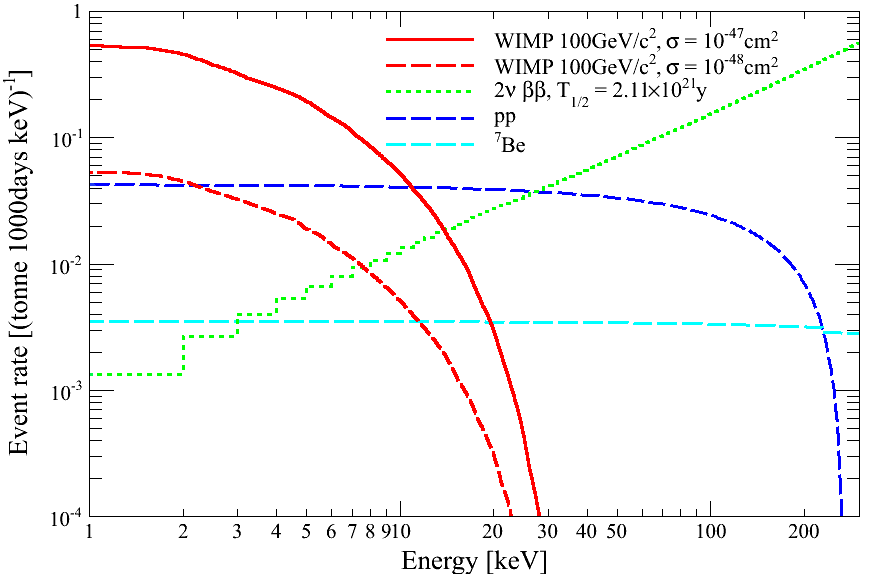}
\includegraphics[width=8.0cm]{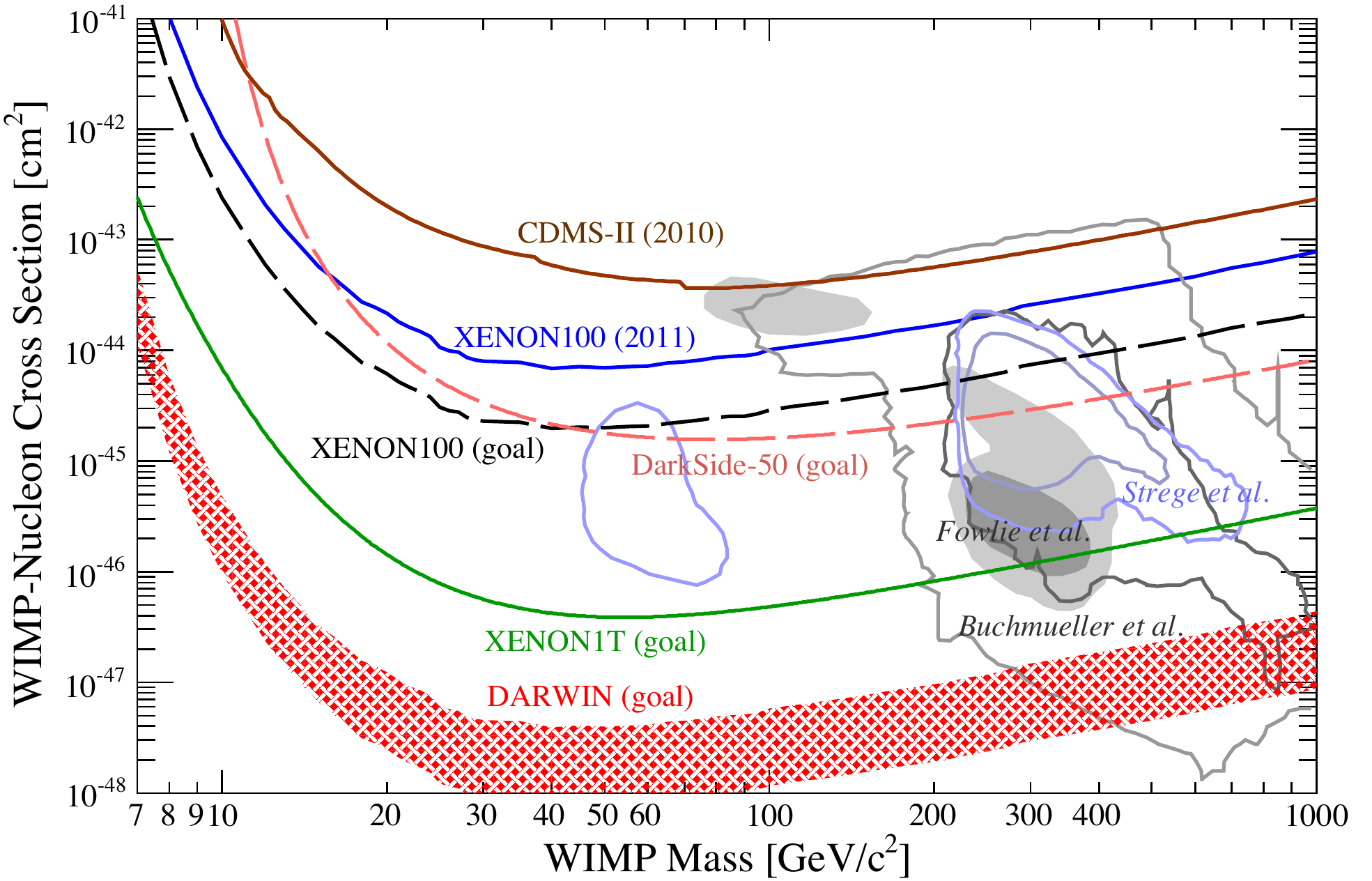}
\caption{\small{(Left): Expected nuclear recoil spectrum from WIMP scatters in LXe for a spin-independent WIMP-nucleon cross section of 10$^{-47}$\,cm$^2$ (red solid) and 10$^{-48}$\,cm$^2$ (red dashed) and a WIMP mass of 100\,GeV/c$^2$ (using the standard halo model as in \cite{complementarity}), along with the differential energy spectrum  for pp (blue) and  $^{7}$Be (cyan) neutrinos, and the electron recoil spectrum from the double beta decay of $^{136}$Xe (green), assuming the natural abundance of 8.9\% and the recently measured half life of  2.1$\times$10$^{21}$\,yr \cite{exo_2011}.  Other assumptions are: 99.5\% discrimination of electronic recoils, 50\% acceptance of nuclear recoils, 80\% flat analysis cuts acceptance. 
(Right): {\small DARWIN's} sensitivity goal for spin-independent WIMP nucleon cross sections, existing limits from XENON100 \cite{xenon100_prl107} and CDMS-II \cite{cdms_science}, future goals and updated theoretical predictions from supersymmetry (closed contours and shaded regions) \cite{fowlie11,buchmueller11,strege11}. }}
\label{fig:diffspectra}
\end{figure}

The left side of figure~\ref{fig:diffspectra}  shows the expected nuclear recoil spectrum from WIMP scatters in xenon together with the background from neutrino-electron elastic scatters of solar neutrinos and from the double beta decay of $^{136}$Xe. The right side shows the aimed sensitivity of {\small DARWIN}, along with existing best upper limits on the WIMP-nucleon cross section, projections for the future and theoretically predicted regions from supersymmetric models.

{\small DARWIN}, which was endorsed in  recent updates to the European and Swiss roadmaps for astroparticle and particle physics \cite{aspera_rm,chipp_rm}, has officially started in 2010.  A rough time schedule is the following: a technical design study is to be ready in spring 2013, leading to a letter of intent and engineering studies towards the proposal of a concrete facility in spring 2014, and a technical design report for detector construction by the end of 2014. The shield and detector construction phase is to start in 2015, commissioning in late 2016 with the start of the first physics run by mid 2017.\\

{\sl Acknowledgements:}
This work is supported through the first {\small ASPERA} common call, by SNF grant 20AS21-129329, by the DoE,  and by the individual institutions participating in the study.

\end{document}